\documentclass[12pt]{iopart}
\usepackage{iopams}

\begin{document}

\title{A new approach to bulk viscosity in strange quark matter at
high densities}

\author{Shou-wan Chen$^1$, Hui Dong$^2$, and Qun Wang$^1$}

\address{$^1$ Department of Modern Physics, University of Science and Technology of China, Hefei 230026, P.R.China}
\address{$^2$ School of Physics, Shandong University, Jinan 250100, P.R.China}
\eads{\mailto{windwan@mail.ustc.edu.cn}, \mailto{hdong@sdu.edu.cn}, \mailto{qunwang@ustc.edu.cn}}


\begin{abstract}
A new method is proposed to compute the bulk viscosity in strange
quark matter at high densities. Using the method it is
straightforward to prove that the bulk viscosity is positive
definite, which is not so easy to accomplish in other approaches
especially for multi-component fluids like strange quark matter with
light up and down quarks and massive strange quarks.
\end{abstract}

\maketitle

There are a lot of works in literature concerning the bulk viscosity
for quark and nuclear matter in compact stars. Some are based on the
Urca and non-leptonic weak processes for nucleons and hyperons
\cite{Langer:1969,Jones:1971,Sawyer:1989dp,Haensel:1992,Prakash:1992,Haensel:2000vz,Haensel:2001mw,Jones:2001ie,Jones:2001ya,Haensel:2001em,Lindblom:2001hd,vanDalen:2003uy,Nayyar:2005th,Chatterjee:2006hy,Chatterjee:2007qs,Gusakov:2007px,Chatterjee:2007iw,Chatterjee:2007ka}.
Others are about Urca processes and $d$-$s$ transitions for unpaired
quark matter
\cite{Wang:1985tg,Sawyer:1989uy,Madsen:1992sx,Dai:1996fe,Anand:1999bj,Xiaoping:2004wc,Dong:2007mb}
and some color superconducting phases
\cite{Sa'd:2006qv,Alford:2006gy,Sa'd:2007ud}. For a brief review,
see e.g. Ref. \cite{Dong:2007ax}. In the CFL phase the bulk
viscosity comes from the lightest degrees of freedom, e.g. kaons,
superfluid $H$ modes etc., which are no longer quarks
\cite{Alford:2007rw,Manuel:2007pz,Alford:2008pb}. In this note, we
will only consider strange quark matter in normal state, for recent
reviews of strange quark matter in stars, see e.g.
\cite{SchaffnerBielich:2004ch,SchaffnerBielich:2007yq}. It is
believed that in the regime of high density far above the saturation
one of nuclear matter, the degrees of freedom become quarks and
gluons. At weak coupling limit for quark matter one can make
reliable predictions based on perturbative quantum chromodynamics.
The bulk viscosity in compact star environment is partly due to
damping mechanism for external perturbations which can lead to
instabilities such as r-mode instability
\cite{Andersson:1997xt,Friedman:1997uh,Lindblom:1998wf,Lai:2001jt,Mannarelli:2008je,
Fu:2008bu}. The stars can drastically spin down by gravitational
radiation when the bulk viscosity is insufficient.

Bulk viscosity arises from energy dissipation due to the change of fluid volume.
The fluid will deviate from thermal and/or chemical equilibrium when it is compressed or rarefied.
The physical pressure (i.e. the instantaneous one) will differ from its equilibrium
value given by the equation of state at the same energy density.
Energy has to be consumed to re-equilibrate the system.
The bulk viscosity, denoted by $\zeta$, is defined through
the rate of energy dissipation,
\begin{equation}
\left\langle \dot{\varepsilon}_{\mathrm{diss}}\right\rangle
=\frac{\zeta}{\tau}\int_{0}^{\tau}(\nabla\cdot\mathbf{v})^{2}dt
=-\frac{n}{\tau}\int_{0}^{\tau}P\dot{V}dt\;,
\end{equation}
where $\tau=2\pi/\omega$ represents the pulsation period, $\nabla\cdot\mathbf{v}$
the divergence of fluid velocity, $n$ the number density, $P$ the
physical pressure, and $V=1/n$ the specific volume.

The bulk viscosity can be defined in an alternative way in terms of complex variables.
The deviation of the physical pressure,
$P$, from the corresponding equilibrium one, $P_{\mathrm{eq}}$, can
be related to the bulk viscosity in the form,
\begin{equation}
\delta P=P-P_{\mathrm{eq}}=-\zeta_{\mathrm{c}}\nabla\cdot\mathbf{v}\;,
\end{equation}
where $\zeta_{\mathrm{c}}$ is complex. The physical bulk viscosity is then
given by the real part $\Re(\zeta_{\mathrm{c}})$.

There are two characteristic time scales related to bulk viscosity,
one corresponds to perturbation and the other to relaxation. The
dominant contribution to dissipative energy loss comes from the
re-equilibration whose time scale is comparable to that of external
perturbations. For example, in strange quark matter which consists
of up and down quarks which are massless, strange quarks which are
massive and electrons, the time scale of the weak interaction in the
system is comparable to the typical one for pulsations induced by
external perturbations. To calculate the bulk viscosity, we can only
consider weak processes.

In literature about the bulk viscosity for quark or nuclear matter,
the simple fluid picture is always implied. In the real world,
however, the strange quark mass is not negligible. For those
external perturbations which do not distinguish particle species,
the pressures felt by different ingredients are the same. This would
lead to equal momentum variations but different velocity ones due to
different masses. Although the time scale for strong interaction
through which the quarks of different flavors reach thermal
equilibrium is much shorter than that of weak interaction
responsible for the bulk viscosity, there are no constraints about
fluid velocities of different species either from hydrodynamic
equations or from particle distribution functions, because the fluid
velocity is the thermal average of particle microscopic velocities,
which vary widely for particles with different masses even when they
have the same temperature. Especially the divergences of the fluid
velocities for particles with different masses are not necessary the
same due to the difference in equations of state, since we have
$\nabla\cdot\mathbf{v}=-\frac{1}{\epsilon +
P}\frac{\partial\epsilon}{\partial t}$  from the fluid equation in
the local co-moving frame, where the energy density $\epsilon$ and
the pressure $P$ depend on masses. Therefore there is no strong
reason that strange quark matter should be regarded as a simple
fluid if the strange quark mass is large compared to temperature and
chemical potential, i.e. the simple fluid picture for strange quark
matter is just an approximation when the strange quark mass is very
large. A natural question to ask is: whether conventional approaches
in literature which are successfully applied to the simple fluid
picture are still valid for multi-component fluids.

Based on above arguments we extended conventional approaches to
strange quark matter as a multi-component fluid. We obtained general
constraints on fluid velocity divergences for different particle
species enforced by baryon number conservation and electric charge
neutrality. We also found a new oscillation pattern which gives a
new solution to the bulk viscosity in the multi-component fluid
satisfying these constraints. The resulting bulk viscosity is about
one order of magnitude larger than those from conventional
approaches. Our further investigations indicate that the positivity
of the results in our previous paper \cite{Dong:2007mb} can not always be guaranteed
for some cases although allowed by those constraints. In this note
we propose an alternative way of computing the bulk viscosity of
strange quark matter as a multi-component fluid in terms of entropy
production, where the bulk viscosity is definitely positive. The
organization of the note is as follows. First we summarize the
formalism to calculate the bulk viscosity for strange quark matter.
Then we introduce our new approach to bulk viscosity by relating the
energy dissipation to entropy production.

In strange quark matter the weak processes via the exchange of
$W^{\pm}$ bosons relevant to bulk viscosity are
\begin{eqnarray}
1.\qquad u+d & \leftrightarrow & u+s\;,\;\;(\mathrm{d\leftrightarrow s\; transition}),\nonumber \\
2.\qquad u+e & \leftrightarrow & d+\nu\;,\;\;(\mathrm{Urca\; I}),\nonumber \\
3.\qquad u+e & \leftrightarrow & s+\nu\;,\;\;(\mathrm{Urca\; II}).
\label{eq:coupled-p}
\end{eqnarray}
Note that the reverse reactions in the second and third lines (Urca
I and II) are not real ones but represent $d/s\to u+e+\bar{\nu}$. In
thermal and chemical equilibrium all chemical potentials satisfy,
\begin{eqnarray}
\mu_{d}& = & \mu_{s}\;,\nonumber \\
\mu_{u}+\mu_{e} & = & \mu_{d}\;,\nonumber \\
\mu_{u}+\mu_{e} & = & \mu_{s}\;.\label{eq:equilibrium}
\end{eqnarray}
Considering there is no accumulation of neutrinos in the system, we
have set $\mu_{\nu/\bar{\nu}}=0$ because neutrino mean free paths
are much longer than typical radii of compact stars. The local
charge neutrality also plays an important role,
\begin{equation} \sum_{i=u,d,s}Q_{i}n_{i}-n_{e}=0\;,
\label{eq:charge_neutrality}
\end{equation}
where $Q_{i}$ are electric charges in unit of $e$. From Eqs.
(\ref{eq:equilibrium}) and (\ref{eq:charge_neutrality}) the
equilibrium values of chemical potentials and number densities can
be obtained.

Volume oscillations induced by perturbations lead to chemical
non-equilibrium, since the chemical potential is related to the
number density $n_{i}$ for quark species $i$ through the equation of
state. The baryon number density is then
$n_{B}\equiv\sum_{i}n_{i}/3$. We can use the oscillation of the
baryon number density, $\delta n_{B}=\delta n_{B0}e^{i\omega t}$, to
account for the volume oscillation. Hereafter we will keep the
subscript $i$ for quarks and $j$ for quarks and electrons. Due to
local charge neutrality, the density oscillations $\delta n_{j}$ are
not independent from each other. The chemical potential differences
between two sides of reactions are written as
\begin{eqnarray}
\Delta\mu_{1} & \equiv & \mu_{s}-\mu_{d}=\delta\mu_{s}-\delta\mu_{d}\;,\nonumber \\
\Delta\mu_{2} & \equiv & \mu_{d}-\mu_{u}-\mu_{e}=\delta\mu_{d}-\delta\mu_{u}-\delta\mu_{e}\;,\nonumber \\
\Delta\mu_{3} & \equiv &
\mu_{s}-\mu_{u}-\mu_{e}=\delta\mu_{s}-\delta\mu_{u}-\delta\mu_{e}\;.\label{eq:coupled-chem}
\end{eqnarray}
The above equations can be put into more compact form,
$\Delta\mu_{k}=c_{kj}\delta\mu_{j}$. Here $c_{kj}$ ($k=1,2,3$ denote
different reactions) are coefficients in front of chemical
potentials in (\ref{eq:coupled-chem}). The reaction rates are
proportional to these differences of chemical potentials when
$\Delta\mu_{k}\ll T (\ll\mu)$,
\begin{equation}
\Gamma_{k}=\lambda_{k}\Delta\mu_{k}\;(=\lambda_{k}c_{kj}\delta\mu_{j})\;.
\label{eq:Gamma_k}
\end{equation}
The coefficients $\lambda_{k}$ can be obtained from Boltzmann
equation with collision terms.

For each ingredient in strange quark matter continuity equations
read
\begin{equation}
\dot{n}_{j}+n_{j}\boldsymbol{\nabla}\cdot\mathbf{v}_{j}=J_{j}\;.
\label{eq:constitute number continuity}
\end{equation}
The source terms in the right-hand-side can be expressed in a linear
combination of $\Gamma_{k}$,
\begin{equation}
J_{j}=-c_{kj}\Gamma_{k}=-c_{kj}\lambda_{k}c_{kj'}\delta\mu_{j'}\,.
\end{equation}
Note that there is no source term for baryon number since it is
conserved in all processes. So we have
\begin{equation}
\dot{n}_{B}+n_{B}\boldsymbol{\nabla}\cdot\mathbf{v}_{B}=0\;.
\label{eq:baryon number continuity}
\end{equation}

The conservation of the baryon number and the electric charge in all
three reactions lead to relations about sources,
\begin{eqnarray*}
\frac{1}{3}\sum_{i}J_{i} & = & 0\;,\\
\sum_{i}Q_{i}J_{i}-J_{e} & = & 0\;.
\end{eqnarray*}
Substituting above relations into continuity equations
(\ref{eq:constitute number continuity}) and (\ref{eq:baryon number
continuity}), we obtain general constraints for fluid velocity
divergences,
\begin{eqnarray}
n_{B}\boldsymbol{\nabla}\cdot\mathbf{v}_{B} & = &
\sum_{i}\frac{1}{3}n_{i}\boldsymbol{\nabla}\cdot\mathbf{v}_{i}\;,\nonumber \\
n_{e}\boldsymbol{\nabla}\cdot\mathbf{v}_{e} & = &
\sum_{i}Q_{i}n_{i}\boldsymbol{\nabla}\cdot\mathbf{v}_{i}\;.
\label{eq:general constraint}
\end{eqnarray}
To solve $n_{j}$ additional inputs are required. In traditional
treatments $n_{B}\dot{X}_{j}=J_{j}$ is used implicitly, where
$X_{j}\equiv n_{j}/n_{B}$ is the number density fraction of baryons
for particle $j$. It can easily be verified from the general
constraints (\ref{eq:general constraint}) that the usage of
$n_{B}\dot{X}_{j}=J_{j}$ implies $\nabla\cdot\mathbf{v}_{j} =
\nabla\cdot\mathbf{v}_{B}$ for all particle species. This holds only
if the mass differences among different particle species are
negligible. However the strange quark mass $m_{s}$ could be much
larger than light quark ones. So the fluid velocity of the strange
component could differ from that of the light sector. In our
previous paper \cite{Dong:2007mb} we considered a limit case where
$\nabla\cdot\mathbf{v}_{s}=0$ and
$\nabla\cdot\mathbf{v}_{u}=\nabla\cdot\mathbf{v}_{d}$,  which
provides additional conditions to fix $n_{j}$. The numerical results
are about one order of magnitude larger than those from traditional
solutions \cite{Wang:1985tg,Sawyer:1989uy,Madsen:1992sx,Dai:1996fe,Anand:1999bj,Xiaoping:2004wc}.

Now we turn to the issue about the positivity of the bulk viscosity.
We will provide an explicit and general demonstration about the
positivity regardless of whether the system is a simple fluid or 
multi-component fluid. The angle in which we look at the problem is
the possible connection between the energy dissipation and the
entropy production related to bulk viscosity arising from
irreversibility of the periodic compression-expansion process. Note
that in an irreversible process the entropy of the system always
increases.

To observe the entropy change we separate the fluid into small
volume cells and focus on one of them. Since the fluid velocity
divergences for particles with different masses are not identical,
or the corresponding hydrodynamic motions are different, the cell is
not a closed system and has exchange of particles with others. Due
to baryon number conservation we can choose the cell in the way that
the total baryon number in the cell does not change. It is an open
system but with a global control number, so we can call it
quasi-control number system. We assume that the quasi-equilibrium is
reached in the cell, i.e. the system is in full thermodynamic
equilibrium except the chemical one, since the chemical reactions
from weak interaction proceed much slower than strong processes. The
entropy of the cell increases during the chemical re-equilibration.
Neglecting the exchange of particles and heat on the surface of the
whole system with outside, the net change of particles and heat
vanishes. Therefore after one cycle of compression-expansion
process, the only thermodynamic quantity that changes is the
entropy. We also neglect temperature change as in traditional
treatments. The change of internal energy in one period reads
\begin{equation}
U(t+\tau)-U(t)  =
\intop_{\tau}\mathrm{d}t\frac{\mathrm{d}}{\mathrm{d}t}\sum_{l}Ts_{N}^{\mathrm{i}}N_{B}^{l}=
N_{B}\intop_{\tau}\frac{T\mathrm{d}s_{N}^{\mathrm{i}}}{\mathrm{d}t}dt\;,
\end{equation}
where the superscript $l$ represents the local cell, $N_{B}^l$ its
baryon number, and $s_{N}^{\mathrm{i}}$ the entropy per baryon
number due to internal irreversible processes. Using the formula for
entropy production in dissipative hydrodynamics we obtain,
\begin{equation}
\frac{T\mathrm{d}s_{N}^{\mathrm{i}}}{\mathrm{d}t}  =
-\mu_{j}\frac{\mathrm{d}X_{j}^{\mathrm{i}}}{\mathrm{d}t} =
\frac{1}{n_{B}\lambda_{k}(\Delta\mu_{k})^{2}}
\end{equation}
Then the second type of definition for bulk viscosity becomes
\begin{eqnarray}
\frac{\zeta}{n_{B}}\intop_{\tau}(\boldsymbol{\nabla}\cdot\mathbf{v}_{B})^{2}\mathrm{d}t
& = & \intop_{\tau}\frac{1}{n_{B}}\lambda_{k}(\Delta\mu)_{k}^{2}\mathrm{d}t\;.
\label{eq:div and internal fraction rate}
\end{eqnarray}
Finally we get the positive bulk viscosity in the new approach,
\begin{equation}
\zeta=\frac{\intop_{\tau}\lambda_{k}(\Delta\mu)_{k}^{2}
\mathrm{d}t}{\intop_{\tau}(\boldsymbol{\nabla}\cdot\mathbf{v}_{B})^{2}\mathrm{d}t}\;.
\end{equation}
It can be proved that the above formula can reproduce the bulk
viscosity for the simple (single-component) fluids.

In summary, from the connection between the dissipative energy loss
and the entropy production a new approach to calculating bulk
viscosity in strange quark matter has been developed. The positivity
of the bulk viscosity can easily be observed in the approach for
simple and multi-component fluids.

{\bf Acknowledgement}. Q.W. is supported in part by '100 talents'
project of Chinese Academy of Sciences (CAS), by National Natural
Science Foundation of China (NSFC) under the grants 10675109 and
10735040. H.D. is supported by National Natural Science Foundation
of China (NSFC) under the grant 10847149.

\end{document}